\documentclass{webofc}
\usepackage[varg]{txfonts}   
\usepackage[english]{babel}
\usepackage{here}
\usepackage{subfloat}
\usepackage{fontenc}
\usepackage[utf8]{inputenc}
\usepackage[colorlinks=false]{hyperref}
\usepackage[printonlyused]{acronym}
\usepackage{bm}
\usepackage{bbold}
\usepackage{amsmath}
\usepackage{slashed}
\usepackage{amssymb}
\usepackage{textcomp}
\usepackage[usenames]{color}
\usepackage{enumerate}
\usepackage{dcolumn}
\usepackage{longtable}
\usepackage[usenames]{color}
\newcommand{\be}{\begin{equation}}
\newcommand{\ee}{\end{equation}}
\newcommand{\bea}{\begin{eqnarray}}
\newcommand{\eea}{\end{eqnarray}}
\newcommand{\ba}{\begin{array}}
\newcommand{\ea}{\end{array}}
\newcommand{\bi}{\begin{itemize}}
\newcommand{\ei}{\end{itemize}}

\begin{document}
\title{Exploration of In-Medium Hyperon-Nucleon Interactions}
%
%

\author{\firstname{Madhumita} \lastname{Dhar}\inst{1}\fnsep\thanks{\email{madhumita.dhar@cgec.org.in}} \and
       \firstname{Horst} \lastname{Lenske}\inst{2}\fnsep\thanks{\email{horst.lenske@physik.uni-giessen.de} }
}

\institute{ Cooch Behar Government Engineering College, Cooch Behar, West Bengal, India 736170
\and
   Institut f\"{u}r Theoretische Physik, Justus-Liebig-Universit\"at Gie\ss en, D-35392 Gie\ss en, Germany       
          }

\abstract{
The study focuses on exploring the changes in the hyperon-baryon interaction at various nuclear densities. This approach starts by building a vacuum hyperon-nucleon interaction model based on Boson -Exchange maintaining SU(3) flavor symmetry. Bethe-Goldstone equation is then explored to investigate the medium properties over the bare interaction. A detailed investigation of the density dependence revealed clear changes in the low energy parameters with the variation of the medium density shown for different strangeness channels.
}
\maketitle
\section{Introduction}\label{sec:intro}
Nuclear physics has gone through an exciting shade after the discovery of strange particles introducing certain  un-explained phenomenon like the 'hyperon puzzle', charge symmetry breaking, hypertriton lifetime puzzle, possibility of NN$\Lambda$ resonant state and so on \cite{Gal}. All these requires knowledge about hyperon-baryon interaction with a special focus on the medium properties. 

With this motivation, this study focuses on the understanding of the hyperon-nucleon interaction in presence of medium that was developed some time ago at Giessen  \cite{dhar}. For that purpose, a microscopic approach is used by first considering the bare interaction and then gradually adding the medium effect to have a thorough understanding of the subject. The major goal of this work is to study the density effects qualitatively first so that with improved experimental inputs, the results can be quantified to a great extent.

Here in this paper, first the methodology for studying vacuum interaction is discussed followed by in-medium effective interaction. The results for each will be discussed. 

\begin{figure}
	\begin{minipage}{0.45\linewidth}
		\centering
		\includegraphics[height=4.1 cm]{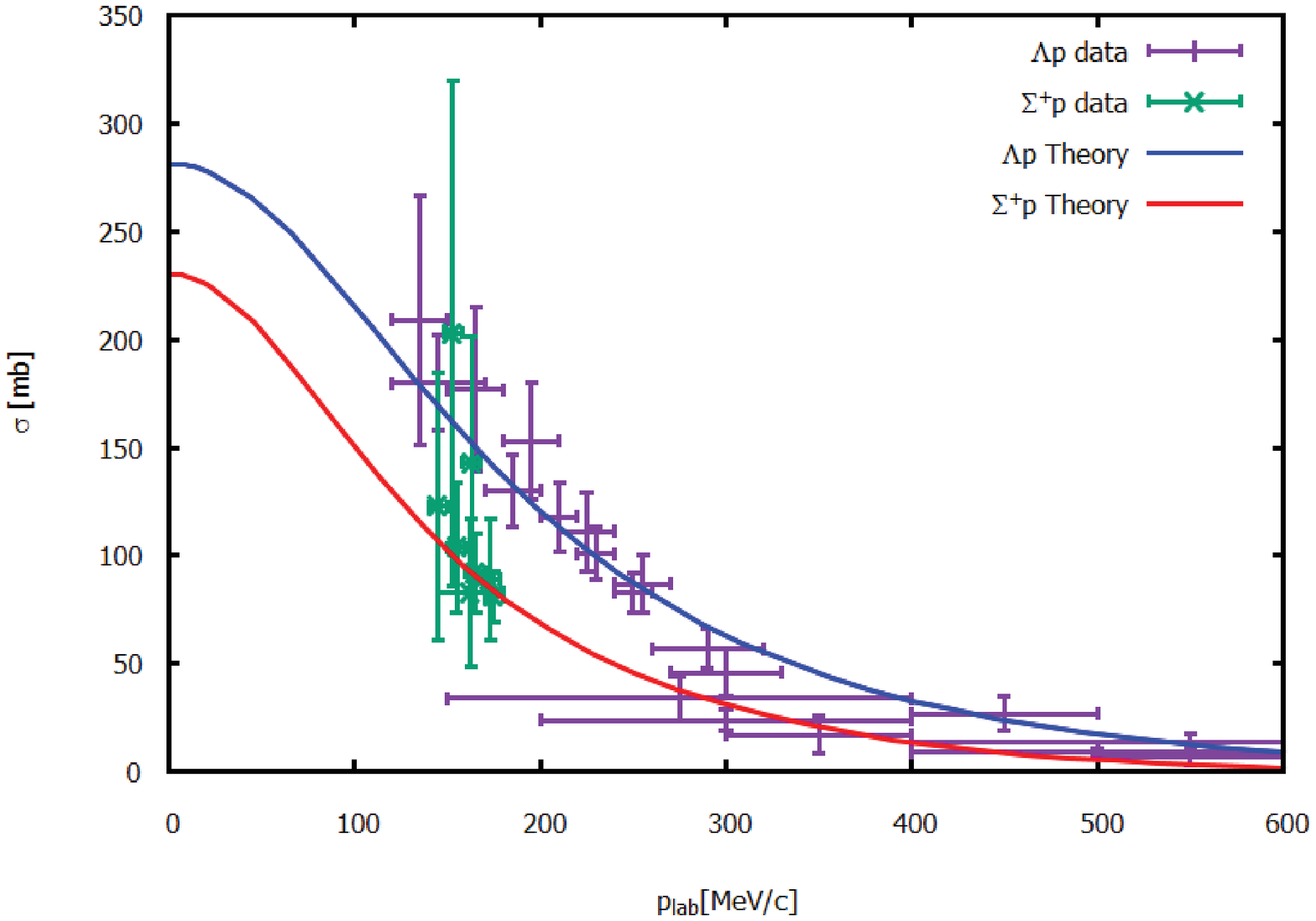}
	\end{minipage}
	\hspace{0.04cm}
	\begin{minipage}{0.45\linewidth}
		\centering
		\includegraphics[height=4.02 cm,width=6.5 cm]{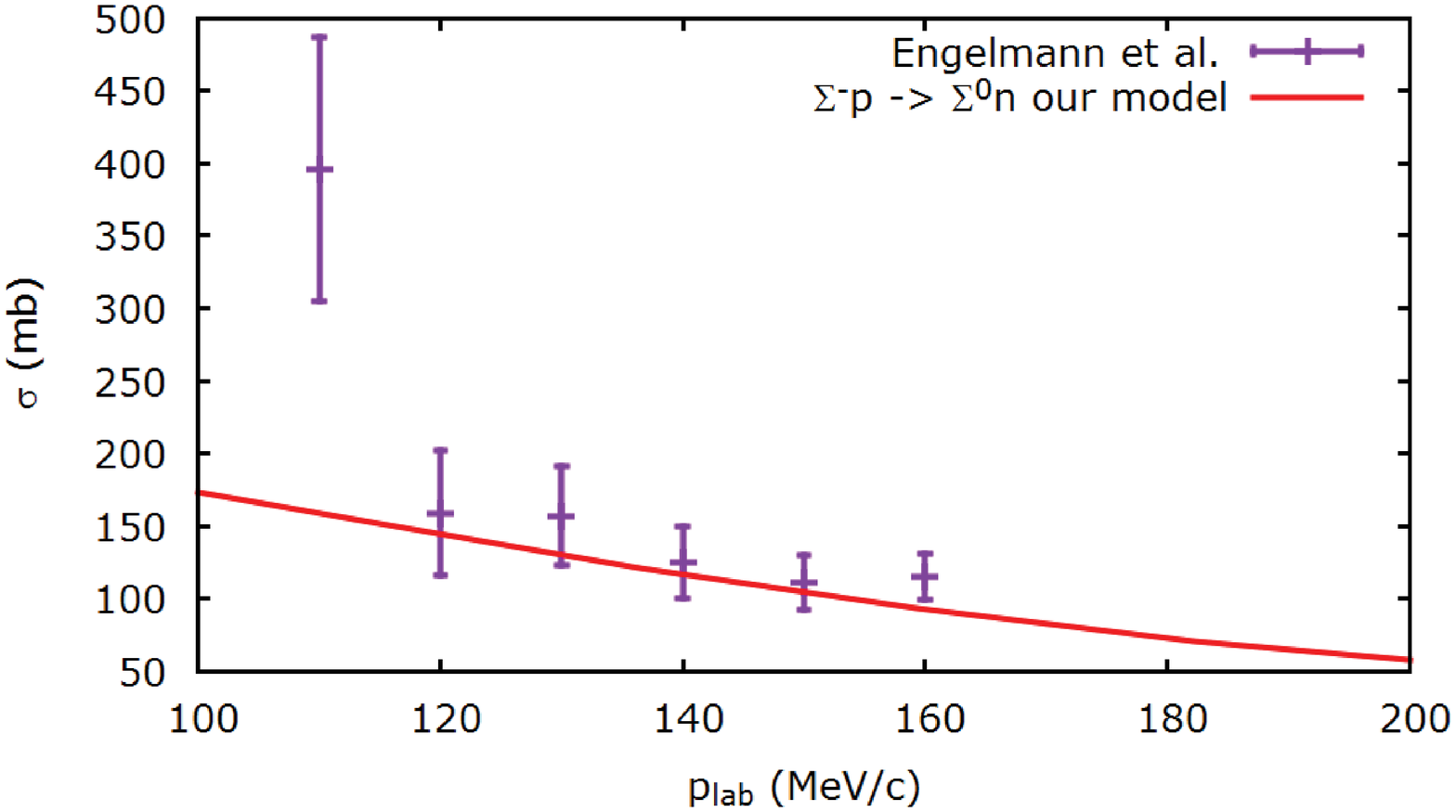}
	\end{minipage}
	\vspace{-0.6cm}
	\caption{ Calculated cross section (theory) for $\Sigma^{+ }p$ (left), $\Lambda p$ (left) and $\Sigma^{- }p$ $\Rightarrow$ $\Sigma^0 n$ (right) channels plotted in comparison to the bubble chamber data \cite{expdata}. }
	\vspace{-0.65cm}
	\label{fig:fit}
\end{figure}

\section{Methodology}
In this work, the meson exchange approach is used to develop a hyperon-nucleon (YN) interaction, applicable to the full baryon and meson octets, for the latter including also the meson singlets. In the past, that strategy has been used extensively, see e.g. \cite{Jue89,Yamamoto:2014jga}, although large uncertainties remain in view of the scarcity of  hyperon-nucleon scattering data. In our approach, the main focus is on in--medium interactions, aiming finally at using hypernuclear data as independent additional constraints on interactions among octet baryons.  

\subsection{Vacuum Hyperon-Baryon Interaction}
Being mainly focused on studying the density effects of the interaction, a revived version of the One- Boson -Exchange potential (OBEP) was developed following SU(3) flavor symmetry to compensate the experimental data set scarcity from theory \cite{deSwart:1963pdg}. The interaction Lagrangian is defined by a superposition of pseudo-scalar (P), scalar (S), and vector (V) SU(3) scalars of the form
\vspace{-0.1cm}
\begin{equation}
 {\mathcal{L}_{int}}=-g_{8}\alpha_{8} Tr([{B,\bar{B}}]~{\phi})+g_{8}(1-\alpha_{8} )Tr(\left\{ {\bar{B},B}\right\}{\phi })+g_1 Tr(B\bar B \phi)\vspace{-0.05cm}\end{equation} with octet coupling constants $g_8, \alpha_8=\frac{F}{F+D}$ and the  singlet coupling $g_1$, $SU(3)_{f}$ invariant baryon and meson matrices B, $\phi$. The mixing of octet and singlet mesons is taken into account by mixing angles $\theta$. 

In order to overcome the uncertainties by the lack of YN-data, here the baryon-meson coupling constants are derived from the elementary g and $\alpha$ couplings by using strictly the fundamental SU(3)-flavor relations \cite{deSwart:1963pdg}. Hence, the three sets of $(g_{8},\alpha_{8},g_{1})_{P,S,V}$ are factors remaining as free parameters. The mixing angles were chosen as the ideal mixing values being found not to be crucial as determining factor if changed from ideal case in affecting the interaction. For $g_{8}^{P}$, the value fixed by pion-nucleon scattering is used. Owing to these, octet strengths ($g_{8}^{S,V})$ for scalar and vector, singlet strengths $(g_{1}^{P,S,V}$) and $ \alpha_{8}^{P,S,V}$ for each three of the meson nonets are fitted in this model.

With this construction and including dipole vertex form factors then the parameters were fitted with the available data set by solving a set of coupled 3-D Lippmann Schwinger equations for the octet T-matrix (Eq. \ref{eq:LippS}) \vspace{-0.25cm}
\begin{equation}
	T (\mathbf{q'},\mathbf{q}) = V (\mathbf{q'},\mathbf{q}) + \int d^{3}k \ V (\mathbf{q'},\mathbf{k})~ G(\mathbf{k}) ~T (\mathbf{k},\mathbf{q})
	\label{eq:LippS}\vspace{-0.09cm}
\end{equation} serving to determine phase shifts, cross sections, and other observabels like low energy parameters with (q, q',k) denoting initial, final and intermediate relative momenta. The OBEP and Green’s function are denoted by V and G, respectively.

\vspace{-0.13cm}
\subsection{Vacuum Interaction Results}
The constructed potential then first fitted to available set of $\Sigma^{+ }p$ and $\Lambda p$ cross section data with $^{1}S_{0}$ partial wave only is at this stage as at this energy sector this is sufficient for preliminary studies. The obtained fit parameter results with a $\chi^2$ fit of 6.68 is shown in Fig. \ref{fig:fit} (left) and the obtained parameter set is given in Table. \ref{table:dataval} (left). As an application, the obtained parameter set were then used to evaluate $\Sigma^{- }p$ $\Rightarrow$ $\Sigma^0 n$ cross section as shown in Fig. [\ref{fig:fit}] (right) to produce a satisfactory output remembering the large error bars of the scattering data. For the present study, only S=-1, -2 results were calculated for few channels as a starting point. 

Thus, having a satisfactory vacuum interaction model, the next step is to apply the medium effect as discussed in next section. 
\vspace{-0.25cm}
\subsection{Medium Effect}\label{sec:med}
To investigate the medium effect, infinite nuclear matter stands as rich laboratory for strong interaction that are crucial for hypernuclei to dense matter to heavy-ion experiments to name a few. In this work, medium effects are incorporated in terms of the Pauli Projector $
Q_{F}= \Theta (k_{1}-k_{F_{1}})\Theta (k_{1}-k_{F_{2}})$ where $k_{F}$ represents nucleon Fermi momentum. $Q_F=1$ in free space. The Bethe-Goldstone equation in momentum space is then given by \vspace{-0.2cm}
\begin{equation}
	T (\mathbf{q'},\mathbf{q}) = V (\mathbf{q'},\mathbf{q}) + \int d^{3}k \ V (\mathbf{q'},\mathbf{k}) ~\mathbf{Q_{F}}(\mathbf{k,k_{F}})~ G(\mathbf{k}) ~T (\mathbf{k},\mathbf{q})
	\label{eq:Bethe_iM}
\end{equation}
\vspace{-1.05cm}
\subsection{In- Medium Results}
\begin{figure}[!htbp]
	\begin{minipage}{0.45\linewidth}
		\centering
		\includegraphics[height=4.27 cm,width=6 cm]{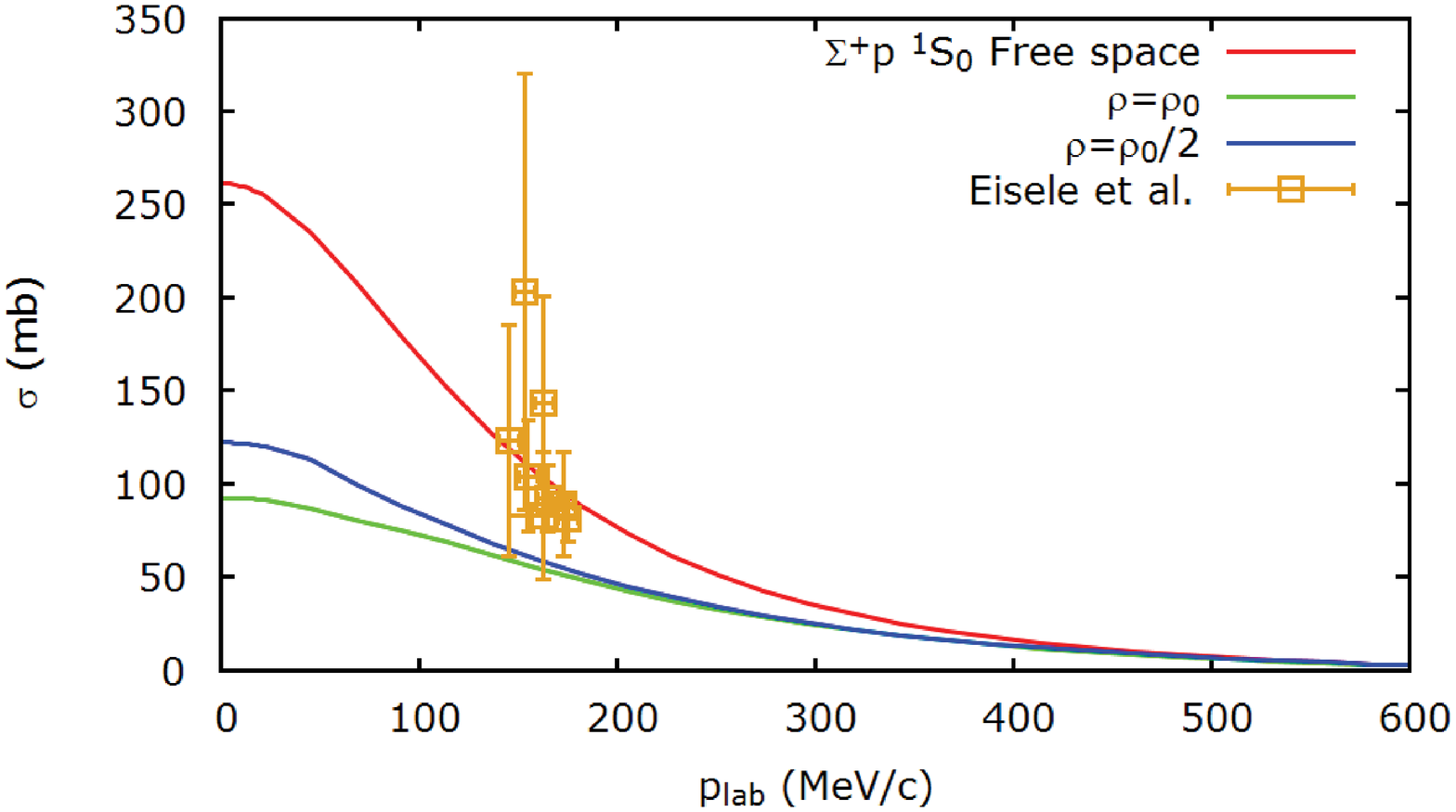}
	\end{minipage}
	\hspace{0.81 cm}
	\vspace{0.17cm}
	\begin{minipage}{0.45\linewidth}
		\centering
		\includegraphics[height=3.8 cm,width=6.2 cm]{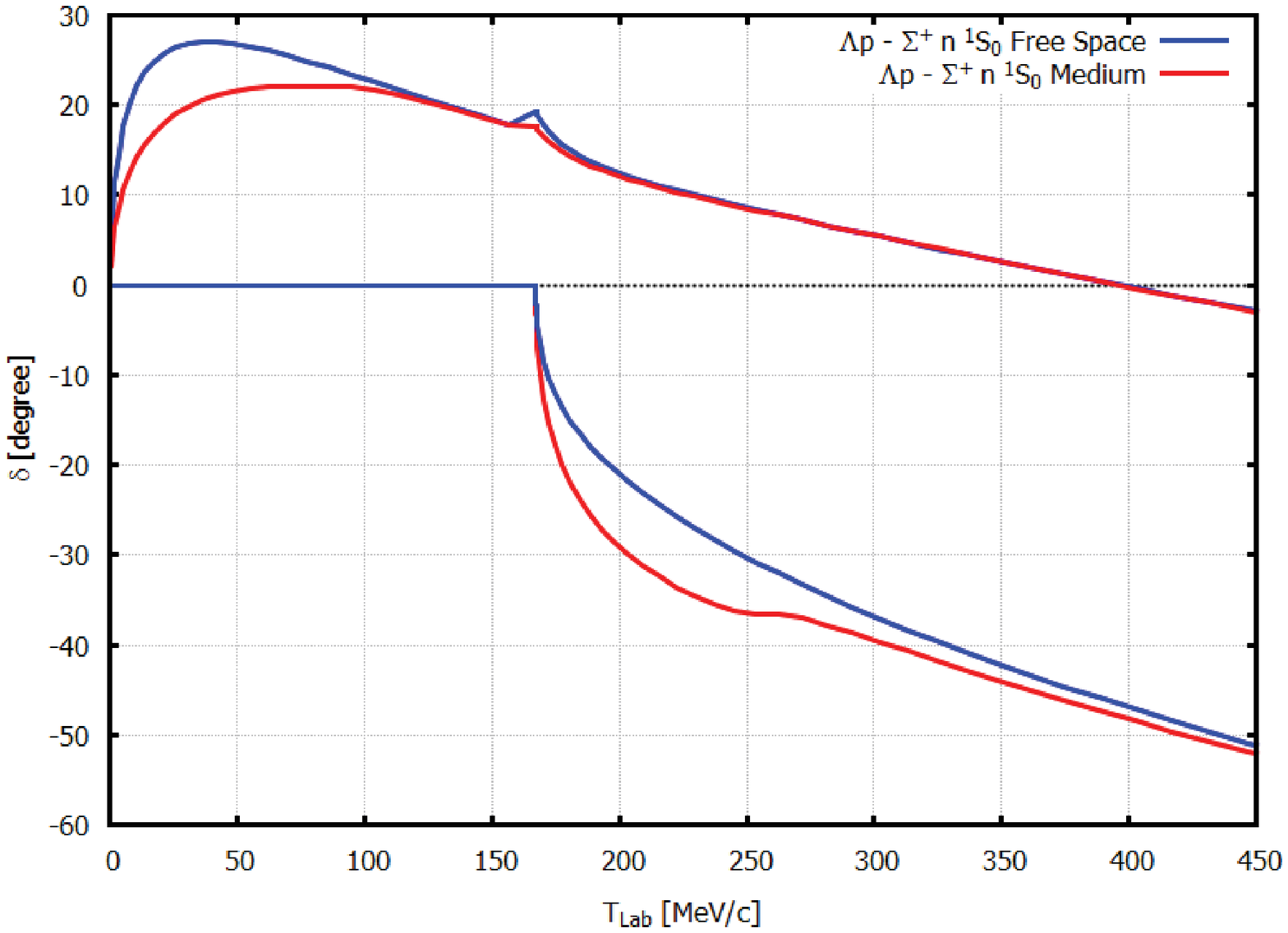}
		\label{fig:cusp}
	\end{minipage}
	\vspace{-0.5cm}
	\caption{ Theoretical integrated $\Sigma^{+} p$ cross-section for nuclear saturation density and half saturation density as a function of laboratory momenta (left) showing significant medium effect on channel opening 'cusp' as can be seen in the phase shift plot (right). }
	\vspace{-0.55cm}
	\label{fig:spXim}
\end{figure}

\begin{figure}[h]
 	\begin{minipage}{0.45\linewidth}
 		\centering
 		\includegraphics[height=4.2 cm,width=5 cm]{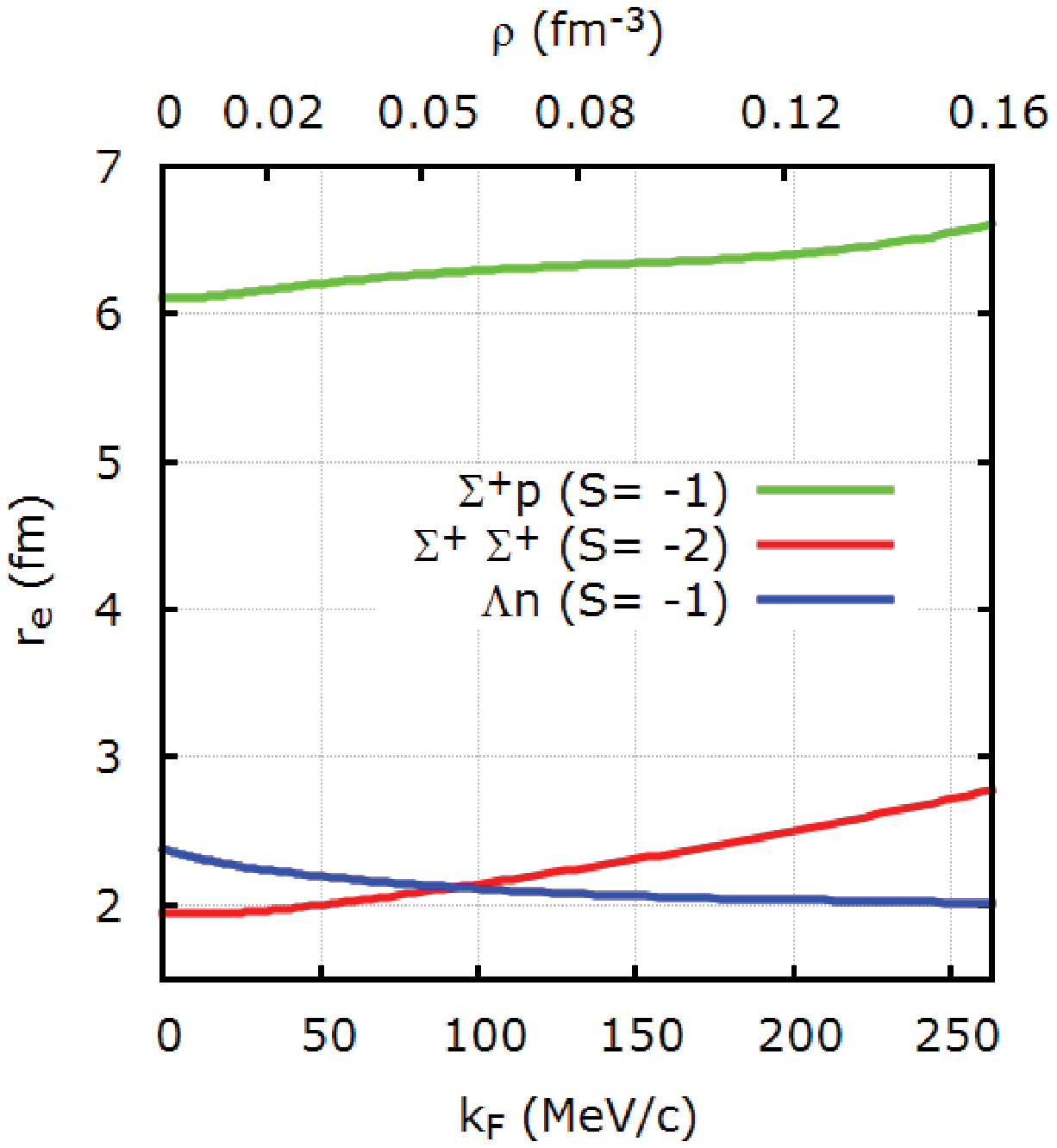}
 		
 	\end{minipage}
 	\hspace{1 cm}
 	\begin{minipage}{0.45\linewidth}
 		\centering
 		\includegraphics[height=4.2 cm,width=5 cm]{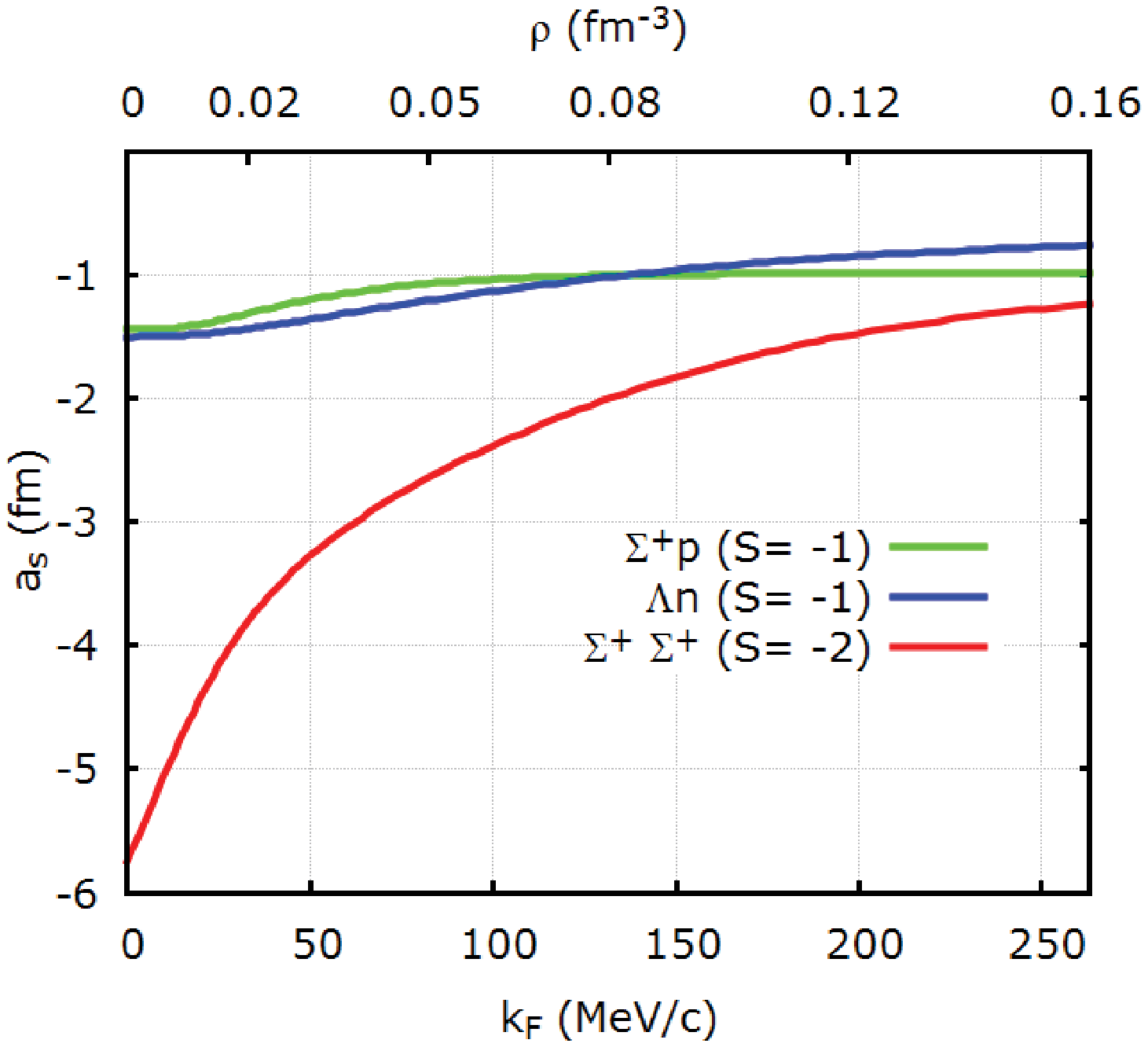}
 		
 	\end{minipage}
 \vspace{-0.2cm}
 	\caption{ Variation of effective range $(r_e)$ (left) and scattering length $(a_s)$ (right)  with nuclear density (top) and Fermi momentum (bottom) revealing channel dependent behavior. }
 	\label{fig:reasim}
 	\vspace{-0.7cm}
 \end{figure}

  When the effect of medium was studied for cross sections, a clear weakening of strength is found as shown in Fig. \ref{fig:spXim} (left). A significant effect of dense medium is seen as suppressing the channel mixing as prominent from the weakening in the sharpness of the 'cusp'- a signature of channel mixing as shown in Fig. \ref{fig:spXim} (right) phase shift plot.
  In order to get direct information for this kind of low energy scattering interaction, effective range parameters are suitable to gain more insights of the interaction. The low energy behavior stands as a convenient  measure for the baryon sector as the core information is limited here (Eq. \ref{eq:eqlow}). \vspace{-0.2cm}
 \begin{equation}
 \lim_{ q\rightarrow 0 } q \cot \delta  \approx -\frac{1}{\textcolor{black}{a_{s}}} + \frac{1}{2} r_{e} q^{2}, ~~ {\textcolor{black}{a_{s} }= \frac{m}{2 \pi \hbar^{2}} \int d^{3}r~ V(r)} 
 \label{eq:eqlow} \vspace{-0.10cm}
\end{equation}
The scattering length ($a_{s}$) and effective range ($r_{e}$) provide information about the nature of interaction. It is important to keep in mind here that for this sector, there are not precise quantitative results yet, hence a qualitative measure can stand as a theoretical prediction.
To explore the density effect on the hyperon interaction, nuclear density was varied and the effect on scattering length and effective range was studied for S=-1 and S=-2 channels. The results obtained are reported in  Fig. \ref{fig:reasim}. Here it shows that the scattering length saturates at saturation density that means at high densities. The calculated values are given in Table. \ref{table:dataval} (right). The obtained results are in-line with the existing predictions as reported by other groups (Extended Soft Core (ESC) \cite{Yamamoto:2014jga}, Juelich \cite{Jue89}, chiral effective field theory ($\chi$EFT) \cite{efts2} models).

\section{Summary and Outlook}
The work reported a revived version of the one-boson exchange potential for hyperon-nucleon interaction primarily aiming at in-medium studies of interactions. As a preliminary step, the vacuum interaction showed good agreement for S=-1 scattering data. With the vacuum input, Bethe Goldstone formalism then successfully describes the dense medium behavior of YN interaction. The investigation also revealed channel dependent behavior as well deserving further exploration.     

As a future step, the approach can be extended for higher strangeness channels as well. The G-matrix formalism can be extended for various in-medium studies of hypernuclei as well to study hypernuclear structure properties in terms of Hartree Fock formalism. As a working qualitative model for hyperon, future promising scattering data availability will be helpful in quantifying the model parameters and thus to have more inputs on the dense matter behavior as well. Inclusion of three--body forces and comparison to the interactions derived independently by the DBHF energy density functional approach in \cite{LenskeDhar:2022} are in preparation.

\vspace{-0.1cm}
 \begin{table}[h]
	\begin{minipage}{0.5\linewidth}
		\begin{tabular}{|c|c|c|c|c||}
			\hline
			& $g_{8}$&  $g_{1}$ &$\alpha_{8}$& $ \theta $ \\\hline\hline
			P & 3.795& 0.1913 & 0.355 &-23 \\ \hline
			S &  1.2274  & 3.5434 &  0.9605 & 37.05 \\ \hline
			V & 1.1566 &3.4431&1.0
			& 35.26 \\ \hline
		\end{tabular}
	\end{minipage}
	\begin{minipage}{0.5\linewidth}
		
		\begin{tabular}{|c|c|c|c|c|}
			\hline
			Channel&$a_{s}^{free}$ & $a_{s}^{sat}$ & $r_{e}^{free}$ & $r_{e}^{sat}$ \\\hline \hline
			$\Lambda n \rightarrow \Lambda n $ & -1.50& -0.76 & 2.34 &2.01 \\ \hline
			$\Sigma^{+} p$ &  -1.44  & -0.86 &  5.18 & 5.34 \\ \hline
			$\Sigma^{+}$ $\Sigma^{+}$ & -5.75 &	-1.23&1.94
			& 2.78\\ \hline
		\end{tabular}
		
	\end{minipage}
	\vspace{-0.3cm}
	\caption{Fitted parameter values of the revived SU(3) OBEP hyperon model (left), Calculated scattering lengths and effective range in free space and nuclear saturation density (right).}
	\vspace{-1.07cm}
	\label{table:dataval}
\end{table}

%

\end{document}